\documentclass[conference,10pt]{IEEEtran}

\usepackage[T1]{fontenc}
\usepackage{graphicx}
\usepackage{cite}
\usepackage{subcaption}
\usepackage{caption}

\usepackage{lipsum}
\usepackage{overpic}
\usepackage{amssymb}
\usepackage{amsmath}
\usepackage{amsthm}
\usepackage{array}
\usepackage{color}
\usepackage{url}

\usepackage{optidef}
\usepackage{times}
\usepackage{fancyhdr,graphicx,amsmath,amssymb}
\usepackage[ruled,vlined]{algorithm2e}

\IEEEoverridecommandlockouts

\theoremstyle{plain}

\newtheorem{lemma}{Lemma}

\newtheorem{corollary}{Corollary}

\newcommand{\vect}[1]{\mathbf{#1}}

\def\diag{\mathrm{diag}}

\def\Ttran{\mbox{\tiny $\mathrm{T}$}}
\def\CN{\mathcal{N}_{\mathbb{C}}} %Complex Gaussian
\def\imagunit{\mathsf{j}} % Imaginary number

\def\sinc{\mathrm{sinc}}

\title{\huge{Intelligent Reconfigurable Surfaces vs. Decode-and-Forward: What is the Impact of Electromagnetic Interference?}\vspace{-0.3cm}}
\author{\IEEEauthorblockN{Andrea de Jesus Torres\textsuperscript{*}, Luca Sanguinetti\textsuperscript{*}, Emil Bj{\"o}rnson\textsuperscript{\textdagger}}
\IEEEauthorblockA{
  \textsuperscript{*}Dipartimento Ingegneria dell'Informazione, University of Pisa, Pisa, Italy\\
  \textsuperscript{\textdagger}Division of Communication Systems, KTH Royal Institute of Technology, Sweden\vspace{-0.5cm}}
  
 }
\begin{document}
	\maketitle
	\vspace{-0.5cm}
\begin{abstract}

This paper considers the use of an intelligent reconfigurable surface (IRS) to aid wireless communication systems. The main goal is to compare this emerging technology with conventional decode-and-forward (DF) relaying. Unlike prior comparisons, we assume that electromagnetic interference (EMI), consisting of incoming waves from external sources, is present at the location where the IRS or DF relay are placed. The analysis, in terms of minimizing the total transmit power, shows that EMI has a strong impact on DF relay-assisted communications, even when the relaying protocol is optimized against EMI. It turns out that IRS-aided communications is more resilient to EMI.
To beat an IRS, we show that the DF relay must use multiple antennas and actively suppress the EMI by beamforming. 
% the comparison between intelligent reflecting surfaces (IRS) are an emerging technology where reflecting metasurfaces can be controlled via software to redirect an impinging waves. IRSs can be used to assist weak wireless communication links, to exploit alternative paths. In past work, it has been shown that an IRS needs lot of elements to compete with a classical alternative, like a decode-and-forward (DF) relay. We want to show how the presence and the distribution of electromagnetic interference (EMI) affects, and possibly reverse, past results. We also propose an algorithm for the DF relay to fight the effects of EMI.
\end{abstract}

\vspace{-.0cm}
\section{Introduction}\vspace{-.0cm}
Intelligent Reflecting Surface (IRS)-aided communication is an emerging topic that is receiving a lot of attention~\cite{direnzo2020,wu2021}. An IRS is a planar array of many reflecting elements with sub-wavelength spacing. Each element can be configured by adjusting its impedance to induce a controllable phase-shift on the incident wave before it is reflected.
By optimizing the phase-shift pattern across the IRS, the reflected wavefront can be shaped as a beam towards the intended receiver~\cite{bjornson2022reconfigurable}. This use case makes the IRS a direct competitor to classical technologies such as amplify-and-forward (AF) \cite{huang2019reconfigurable} and decode-and-forward (DF) relays~\cite{bjorn2020irs}. A detailed comparison between IRS-aided communications and repetition-coded DF relaying was provided in~\cite{bjorn2020irs}. The main conclusion in~\cite{bjorn2020irs} is that high rates and/or large surfaces are needed to beat DF relaying.

The analysis in~\cite{bjorn2020irs} considered only the signal generated by the source and neglected the electromagnetic interference (EMI) or ``extrinsic noise'' (cf.~\cite{nossek2009}). The EMI may arise from a variety of causes, e.g., other (single or multiple) transmitting devices and/or natural background radiation. {In the former case, its strength may typically be from $30$~dB to $10$~dB weaker than the source signal, while, in the latter case, it can even be smaller than the thermal noise power. Despite its strength being highly dependent on the considered wireless environment, we can reasonably assume that (weak or strong) EMI will always be present, unless the communication takes place in an anechoic chamber designed to completely absorb reflections of uncontrollable electromagnetic waves.} In~\cite{torres2021}, a physically meaningful model for EMI was provided and used to show that, in a random scattering environment, the EMI may severely impact IRS-aided communications, especially when the IRS is large. When a non-negligible direct link is present, the IRS might even reduce the communication rate. 

Motivated by these observations, in this paper we return to the comparison in~\cite{bjorn2020irs} and assume that EMI is present at the location where the IRS and DF relay are deployed. For a fair comparison, we start assuming that both technologies use a single radio-frequency chain {(the IRS needs it for control purposes \cite{bjornson2022reconfigurable})}. The comparison is carried out assuming that both technologies are either optimized or not against EMI. It turns out that the IRS outperforms the DF relay, thanks to its spatially filtering capabilities, which increase as the number of IRS elements grows large. We show that the performance gap can only be reduced if multiple antennas are used at the DF relay together with interference-suppressing signal processing.

\begin{figure}\vspace{-0.5cm}
\centering
	\begin{overpic}[width=.95\columnwidth]{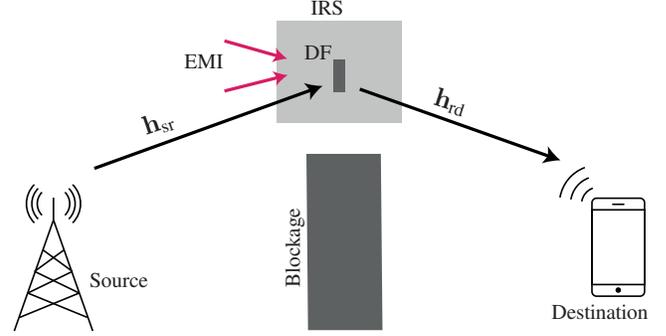}
	\put(20,31){\rotatebox{19}{{$\vect{h}_{\text{sr}}$}}}
	\put(66,36){\rotatebox{-18}{{$\vect{h}_{\text{rd}}$}}}
	\put(47,50){\rotatebox{0}{\footnotesize{IRS}}}
	\put(46,43){\rotatebox{0}{\footnotesize{DF}}}
	\put(43,7){\rotatebox{90}{\footnotesize{Blockage}}}
	
	\put(12,7){\rotatebox{0}{\footnotesize{Source}}}
	\put(85,2){\rotatebox{0}{\footnotesize{Destination}}}
	
	\put(27,41.5){\rotatebox{0}{\footnotesize{{EMI}}}}
	
	\end{overpic}

	\caption{The BS is linked to the UE only through a one-hop IRS/DF assisted link. The direct link is blocked.}\vspace{-0.5cm}
	\label{fig:1Hop}
\end{figure}

% \begin{figure}
% 	\begin{overpic}[width=.9\columnwidth]{2Hop.eps}
% 	\put(10,28){\rotatebox{49}{{$\vect{h}_{\text{BR}}$}}}
% 	\put(48,41){\rotatebox{0}{{$\vect{h}_{\text{rr}}$}}}
% 	\put(81,38){\rotatebox{-59}{{$\vect{H}_{\text{RU}}$}}}
% 	\put(20,49){\rotatebox{0}{\footnotesize{IRS}}}
% 	\put(20,43){\rotatebox{0}{\footnotesize{DF}}}
% 	\put(43,11){\rotatebox{90}{\footnotesize{Blockage}}}
	
% 	\put(12,7){\rotatebox{0}{\footnotesize{BS}}}
% 	\put(85,10){\rotatebox{0}{\footnotesize{UE}}}
	
% 	\put(22,21){\rotatebox{0}{\footnotesize{{EMI}}}}
% 	\put(73,21){\rotatebox{0}{\footnotesize{{EMI}}}}
	
% 	\end{overpic}
	
% 	\caption{The BS is linked to the UE only through a two-hop IRS/DF assisted link. The direct link from BS to UE is fully obstructed by a blockage, as well as the links from the first IRS/DF toward the UE and from the second IRS/DF toward the BS.}
% 	\label{fig:2Hop}
% \end{figure}
% 
% \begin{figure}
% 	\begin{overpic}[width=.9\columnwidth]{Setup.eps}
    
% 	\put(20,27){\rotatebox{0}{\footnotesize{\textcolor{black}{IRS 1}}}}
% 	\put(93,27){\rotatebox{0}{\footnotesize{\textcolor{black}{IRS 2}}}}
	
% 	\put(22,3){\rotatebox{0}{{\textcolor{black}{x}}}}
% 	\put(3,23){\rotatebox{0}{{\textcolor{black}{y}}}}
% 	\put(6,-1){\rotatebox{0}{\footnotesize{\textcolor{black}{Source}}}}
% 	\put(73,7){\rotatebox{0}{\footnotesize{\textcolor{black}{Destination}}}}
% 	\end{overpic}
	
% 	\caption{Caption}
%     \label{fig:simul_setup}
% \end{figure}

\vspace{-.1cm}
\section{IRS-Assisted Communications With EMI}\vspace{-.1cm}
Consider the system depicted in Fig.~\ref{fig:1Hop} in which a single-antenna BS communicates with a single-antenna receiver. We assume the direct link is blocked and, thus, the transmission is assisted by an IRS equipped with $N$ reconfigurable elements that are deployed edge-to-edge on a two-dimensional square grid. %The IRS is assumed to be configurable by means of a single antenna (i.e., single radio-frequency chain) communication link. 
The elements have an area $A = (\lambda/2)^2$ and are indexed row-by-row by $n=1,\ldots,N$. We use $\vect{u}_n$ to indicate the location of the $n$th element. The IRS configuration is determined by the diagonal matrix $\boldsymbol{\Phi} =  \diag(e^{\imagunit \phi_1},\ldots,e^{\imagunit \phi_N})$ where $\phi_1,\ldots,\phi_N\in[0,2\pi)$ are the controllable phase-shift variables. The channel vector between the BS and IRS is $\vect{h}_{\text{sr}} \in \mathbb{C}^N$  and the channel vector between the IRS and receiver is $\vect{h}_{\text{rd}} \in \mathbb{C}^N$. We assume deterministic propagation conditions with the source and destination being in the far-field of the IRS and a single line-of-sight (LoS) path. 
We call $\{\varphi_{\text{sr}},\theta_{\text{sr}}\}$ and $\{\varphi_{\text{rd}},\theta_{\text{rd}}\}$ the azimuth and elevation angles of the single path for the two links. In this case, we have that $[{\bf h}_{\rm{sr}}]_n = \sqrt{\beta_{\rm{sr}}}e^{\imagunit \vect{k}(\varphi_{\rm{sr}}, \theta_{\rm{sr}})^{\Ttran}\vect{u}_n}$ and $[{\bf h}_{\rm{rd}}]_n = \sqrt{\beta_{\rm{rd}}}e^{\imagunit \vect{k}(\varphi_{\rm{rd}}, \theta_{\rm{rd}})^{\Ttran}\vect{u}_n}$
where $\beta_{\rm{sr}}=|h_{{\rm{sr}},n}|^2$ and $\beta_{\rm{rd}}=|h_{{\rm{rd}},n}|^2$ are the channel gains.

\subsection{Signal model}

Unlike most of the existing literature, we assume that EMI is present (at the location where the IRS is placed), generated by source-independent emitters. Following~\cite{torres2021}, the EMI vector $\vect{n} \in \mathbb{C}^N$ at the IRS is modelled as $\vect{n} \sim \CN(\vect{0},\sigma_{\rm{emi}}^2\vect{R}) $ where 
%the $(n,m)$th element of $\vect{R}$ is given by
\begin{align}
\left[\vect{R}\right]_{n,m} =\int_{0}^{\pi} \hspace{-2mm} \int_{-\pi}^{0} e^{\imagunit\vect{k}(\varphi,\theta)^{\Ttran}(\vect{u}_n-\vect{u}_m)} f(\varphi,\theta) d \varphi d \theta.  \label{eq:R1_expression}
\end{align} 
Here, $\sigma_{\rm{emi}}^2f(\varphi,\theta)$ denotes the EMI's power angular density, $\varphi$ is the azimuth angle, $\theta$ is the elevation angle and $\vect{k}(\varphi, \theta)= \frac{2\pi}{\lambda}[\cos(\theta) \cos(\varphi), \cos(\theta) \sin(\varphi),\sin(\theta)]^{\Ttran}$ is the wave vector. The above model is valid for any arbitrary $f(\varphi,\theta)$. 
%, but in this paper we will focus on an $f(\varphi,\theta)$ in the form:
% \begin{align}\label{eq:non-isotropic_pdf}
% f(\varphi,\theta)  = \frac{c}{2 \pi \bar\sigma_{\varphi} \bar\sigma_{\theta}} e^{-\frac{(\varphi - \bar\varphi)^2}{2 \bar\sigma_{\varphi}^2}}e^{-\frac{(\theta - \bar\theta)^2}{2 \bar\sigma_{\theta}^2}} \cos (\theta)
% \end{align}
% where $c$ is a constant such that $\int_{0}^{\pi} \hspace{-1mm} \int_{-\pi}^{0} f_i(\varphi,\theta) d\varphi d \theta = 1$. 
%
In the presence of isotropic EMI,~\eqref{eq:R1_expression} reduces to \cite[Prop. 1]{Bjornson21}
\begin{align}\label{eq:correlation_matrix}
[\vect{R}^{{\rm{iso}}}]_{n,m} =\sinc \left(\frac{2\|\vect{u}_{n}-\vect{u}_{m}\|}{\lambda}\right)
\end{align}
where $\| \cdot \|$ denotes the Euclidean norm. 
%This condition is met for $ \bar\sigma_{\varphi}^2 =\bar\sigma_{\theta}^2\rightarrow \infty$ in Eq.~\eqref{eq:non-isotropic_pdf}.
%
With EMI, the signal received at the destination is~\cite[Eq.~(3)]{torres2021} 
\begin{align}
	y_d = 	\vect{h}_{\rm{rd}}^T \vect{\Phi}\left(\vect{h}_{\rm{sr}}\sqrt{p}s + \vect{n}\right) + w
\end{align}
where $p$ denotes the transmit power at the source and $w \sim \CN(0,\sigma^2)$ is the thermal noise at the destination. 

\subsection{Performance analysis}
Under the assumption of deterministic channels, the information rate (bit/s/Hz) of the end-to-end channel is 
%The signal-to-noise ratio (SNR) is
%\begin{align}
%	\rm{SNR}=\frac{p|\vect{h}_{\rm{rd}}^T \vect{\Phi}\vect{h}_{\rm{sr}}|^2}{\sigma_{\rm{emi}}^2|\vect{h}_{\rm{rd}}^T \vect{\Phi}\vect{R}^{1/2}|^2 + \sigma^2}.
%	\label{eq:SNR}
%\end{align}
\begin{equation}
	R_{\rm{IRS}} = \log_2\left(1 + \frac{p|\vect{h}_{\rm{rd}}^T \vect{\Phi}\vect{h}_{\rm{sr}}|^2}{\sigma_{\rm{emi}}^2\|\vect{h}_{\rm{rd}}^T \vect{\Phi}\vect{R}^{1/2}\|^2 + \sigma^2}\right).
	\label{eq:R1IRS}
\end{equation}
The optimal phase-configuration in the presence of only thermal noise is $\phi_n = -\arg([{\bf h}_{\rm{sr}}]_{n} [{\bf h}_{\rm{rd}}]_{n})$, which maximizes the numerator.
%With the optimal phase-configuration against thermal noise only, the phases of $\vect{\Phi}$ are chosen as $\phi_n = -\arg([{\bf h}_{\rm{sr}}]_{n} [{\bf h}_{\rm{rd}}]_{n})$.
%and~\eqref{eq:R1IRS} becomes
%\begin{align}
%	R_{\rm{IRS}} = \log_2\left(1 + \frac{p\beta_{\rm{rd}}\beta_{\rm{sr}}N^2}{\sigma_{\rm{emi}}^2|\vect{h}_{\rm{rd}}^T \vect{\Phi}^\star\vect{R}^{1/2}|^2 + \sigma^2}\right).
%	\label{eq:R1IRS_opt}
%\end{align}
In the presence of EMI, better performance can be achieved by tuning the IRS to the EMI statistics, i.e., knowledge of $\vect{R}$. In~\cite[Sec. IV]{torres2021}, a heuristic algorithm is designed based on the projected gradient descent to compute a suboptimal $\vect{\Phi}$ that achieves higher rate. 

If a particular rate $\bar R$ is required at the destination, the rate expression in~\eqref{eq:R1IRS} can be used to compute the transmit power that is required with and without the optimization against EMI.

\begin{lemma}
To achieve a particular rate $\bar R$ in the presence of EMI, the IRS-aided communications require the power
\begin{align}\label{eq:opt_power_ris}
    p = \left(2^{\bar R} - 1\right)\frac{\sigma_{\rm{emi}}^2\|\vect{h}_{\rm{rd}}^T \vect{\Phi}\vect{R}^{1/2}\|^2 + \sigma^2}{|\vect{h}_{\rm{rd}}^T \vect{\Phi}\vect{h}_{\rm{sr}}|^2}.
\end{align}
%\begin{align}
% \vect{\Phi}^\star = \begin{cases}
%\phi_n^\star = -\arg([{\bf h}_{\rm{sr}}]_{n} [{\bf h}_{\rm{rd}}]_{n}) $ {\textnormal{if optimized against thermal noise}}$\\
%\;\; \textnormal{Iterative algorithm~\cite{torres2021}.}
%\end{cases}
%\end{align}
%\begin{align}
%\phi_n = -\arg([{\bf h}_{\rm{sr}}]_{n} [{\bf h}_{\rm{rd}}]_{n})
%\end{align}
\end{lemma}
This lemma assumes a fixed $\vect{\Phi}$. The phase-shift variables could be selected as $\phi_n = -\arg([{\bf h}_{\rm{sr}}]_{n} [{\bf h}_{\rm{rd}}]_{n})$ if $ \vect{\Phi}$ is optimized against thermal noise only or  obtained by the iterative algorithm in~\cite{torres2021} if it is tuned based on EMI statistics.

\section{Relay-Assisted Communications With EMI}
In this section, we still consider the system in Fig.~\ref{fig:1Hop} but assume now that the communication is assisted by a half-duplex relay that is deployed exactly in the same position as the IRS. We consider a DF relaying protocol where the transmission is divided into two subsequent phases. We call $0<\tau_1,\tau_2<1$  the fraction of channel uses assigned to the two phases, with $\tau_1 + \tau_2 =1$. The average  transmit power is 
\begin{align}\label{eq:power}
p =\tau_1p_1 + \tau_2p_2  = \tau_1 p_1 + (1 - \tau_1)p_2  
\end{align}
where $p_1$ and $p_2$ are the transmit powers in the first and second phase, respectively. For a given rate requirement, we can select $\tau_1,\tau_2,p_1,p_2$ to minimize $p$. As for the IRS, we assume that the DF is equipped with a single radio-frequency chain, i.e., a single-antenna. This makes the two technologies comparable in terms of complexity. The benefit of having multiple radio-frequency chains at the DF will be investigated in Section~\ref{sec:number_antennas}.

%DF-relay. As in the previous section, both source and destination are single-antenna devices and the direct link between them is completely obstructed by the blockage. The DF-relay has M antennas and deterministic propagation conditions are assumed. The communication strategy for the DF-relay differs from that of the IRS, as, being a digital device, it can not operate in full-duplex mode. The communication is then performed in two subsequent phases: the first from source to DF and the second from DF to destination. Though the time shares allocated for each phase are usually equal, we will also provide results in case of time allocations optimization.

\subsection{Signal model}
Following~\cite{bjorn2020irs}, the signal received at the DF with EMI is
\begin{align}\label{eq:DF_first_stage}
	y_{\rm 1} = h_{\rm sr} \sqrt{p_1} s + n + {w}_{\rm 1}
\end{align}
where ${w}_{\rm 1} \sim \CN({0},\sigma^2)$ is the thermal noise at the DF relay and $n \sim \CN({0},\sigma_{\rm  emi}^2)$ is the EMI with the same variance as earlier. In the second phase, the signal received at the destination is
\begin{align}\label{eq:DF_second_stage}
	y_{\rm 2} = h_{\rm rd} \sqrt{p_2} s + w_2
\end{align}
where ${w}_2\sim \CN(0,\sigma^2)$ is the thermal noise at the destination. 
%By utilizing~\eqref{eq:DF_first_stage} and~\eqref{eq:DF_second_stage} with maximum ratio combining and precoding (i.e., $\vect{g} = \vect{h}_{\rm sr} $ and $\vect{v} = \vect{h}_{\rm rd}^* $), the 
The achievable rate is
\begin{equation}
	R_{\rm DF} = \textrm{min}\left\{ \tau_1\log_2\left(1 + p_1 \alpha_1\right), \tau_2\log_2\left(1 + p_2 \alpha_2\right) \right\}\!\!
	\label{eq:Rdtfs}
\end{equation}  
where $
	\alpha_1 = \frac{\beta_{\rm sr}}{\sigma_{\rm emi}^2 + \sigma^2}$
and $
	\alpha_2 = \frac{\beta_{\rm rd}}{\sigma^2}$
are the effective channel gains in the first and second phase, respectively. 

\subsection{Performance analysis}

In the simplest case of repetition-coded relaying (i.e., $ \tau_1 =  \tau_2 = 1/2$), the maximum achievable rate is~\cite[Eq.~(14)]{bjorn2020irs}
\begin{align}
	R_{\rm DF} = \frac{1}{2}\log_2\left(1 + \frac{2p\beta_{\rm rd}\beta_{\rm  sr}}{\beta_{\rm  sr}\sigma^2 + \beta_{\rm  rd}( \sigma_{\rm  emi}^2 + \sigma^2)}\right)
\end{align}
and is obtained by setting $p_1 \alpha_1 = p_2 \alpha_2$ under the constraint $p = \frac{p_1 + p_2}{2}$, which  follows from \eqref{eq:power} for $\tau_1 =  \tau_2 = 1/2$.
\begin{corollary}[$\!\!\!$\cite{bjorn2020irs}]
To achieve a particular rate $\bar R$ in the presence of EMI, the repetition-coded DF relaying %(i.e., $ \tau_1 =  \tau_2 = 1/2$) 
requires the power\vspace{-2mm}
\begin{align}\label{eq:opt_power_df}
	p = (2^{2\bar{R}}-1)\frac{\beta_{\rm  sr}\sigma^2 + \beta_{\rm  rd}(\sigma_{\rm  emi}^2 + \sigma^2)}{2\beta_{\rm rd}\beta_{\rm  sr}}.
\end{align}
\end{corollary}
A lower transmit power can be achieved by tuning $\tau_1$ (or, equivalently, $\tau_2$ since $\tau_2 + \tau_1 = 1$) and $\{p_1,p_2\}$ according to the propagation conditions. From~\eqref{eq:power}, this requires to solve the following optimization problem:
\begin{align}\label{eq:opt_problem_rate}
\min_{\tau_1,p_1,p_2} \quad {\tau_1 p_1 + (1 - \tau_1)p_2  }  \quad \quad {\text {s.t.}}  \quad R_{\rm DF} \ge \bar R.
\end{align}
The solution to~\eqref{eq:opt_problem_rate} is obtained by Algorithm 1. The outer while-loop performs a bisection search for the minimum power $p$. At each iteration $n$, the inner stage performs an \emph{exhaustive search} to solve
\begin{align}\label{eq:opt_problem}
&\max_{\tau_1,p_1,p_2} \; \textrm{min}\left\{ \tau_1\log_2\left(1 + p_1 \alpha_1\right), \tau_2\log_2\left(1 + p_2 \alpha_2\right) \right\}   \!\!\\ &\quad {\text {s.t.}} \quad \; \tau_1 p_1 + (1 - \tau_1)p_2  \le p_{[n]}\notag
\end{align}
which returns the maximum achievable rate $R^\star_{{\rm DF}}(p_{[n]})$ for a total budget power given by $p_{[n]}$.

\begin{algorithm}
\SetKwRepeat{Do}{do}{while}%
\SetKwInOut{Input}{Input}
\SetKwInOut{Output}{Output}
\DontPrintSemicolon
\Input{$\bar R$, $\alpha_1$, $\alpha_2$, $\epsilon$}
\Output{$\tau_1,p_1,p_2$}
\tcc{Initialization}
$p^{\mathrm{lower}} \leftarrow 0$, $p^{\mathrm{upper}} \leftarrow 10^5$, $\epsilon \leftarrow 10^{-6}$\;
Solve~\eqref{eq:opt_problem_rate} to obtain $R^\star_{{\rm DF}}(p^{\mathrm{lower}})$ and $R^\star_{{\rm DF}}(p^{\mathrm{upper}})$\;

\tcc{Bisection method}
\Do{$(R^\star_{{\rm DF}}(p_{[n]})-\bar R)/\bar R \geq \epsilon$}{
	$p_{[n]} \leftarrow \frac{p^{\mathrm{lower}}+p^{\mathrm{upper}}}{2}$\;
	Compute $R^\star_{{\rm DF}}(p_{[n]})$ by solving~\eqref{eq:opt_problem_rate}\;
    \uIf{$(R^\star_{{\rm DF}}(p_{[n]})-\bar R) (R^\star_{{\rm DF}}(p^{\mathrm{upper}})-\bar R) \leq 0$}{
      $p^{\mathrm{lower}} \leftarrow p_{[n]}$\; 
      Compute $R^\star_{{\rm DF}}(p^{\mathrm{lower}})$ by solving~\eqref{eq:opt_problem_rate}\;
    }
    \Else{
      $p^{\mathrm{upper}} \leftarrow p_{[n]}$\;
      Compute $R^\star_{{\rm DF}}(p^{\mathrm{upper}})$ by solving~\eqref{eq:opt_problem_rate}\;
    }
}
\caption{Bisection algorithm for solving~\eqref{eq:opt_problem}.}
\label{alg:max-min-fairness}
\end{algorithm}

\section{Numerical Comparisons}

\begin{figure}[t!]\vspace{-0.5cm}
\centering
	\begin{overpic}[width=.9\columnwidth]{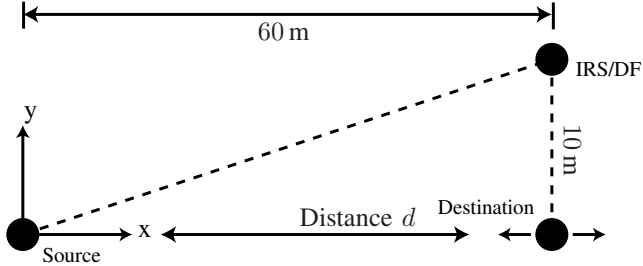}
    
    \put(42,36){\rotatebox{0}{$60$\,m}}
	\put(93,22){\rotatebox{-90}{$10$\,m}}
	\put(95,30){\rotatebox{0}{\footnotesize{\textcolor{black}{IRS/DF}}}}
	\put(22,3){\rotatebox{0}{{\textcolor{black}{x}}}}
	\put(3,23){\rotatebox{0}{{\textcolor{black}{y}}}}
	\put(6,-1){\rotatebox{0}{\footnotesize{\textcolor{black}{Source}}}}
	\put(72,7){\rotatebox{0}{\footnotesize{\textcolor{black}{Destination}}}}
    \put(49,5){\rotatebox{0}{Distance $d$}}
	\end{overpic}
	
	\caption{System setup with the source, destination and IRS/DF placed in the $xy$-plan.}\vspace{-.5cm}
    \label{fig:simul_setup}
\end{figure}

We will now compare two different technologies numerically. We consider the setup depicted in Fig.~\ref{fig:simul_setup} where the coordinates of the elements are: $\rm{Source}=[0,~0,~0]$\,m; $\rm{IRS/DF}=[60,~10,~0]$\,m and ${\rm Destination}=[d,~0,~0]$\,m. 

We model the channel gain according to the 3GPP Urban Micro (UMi) from~\cite[Table B.1.2.1-1]{3gpp} and the carrier frequency is set at $3$\,GHz, corresponding to $\lambda = 0.1$\,m. The shadow fading is neglected to obtain a reproducible deterministic channel. The bandwidth is  $B=10$\,MHz while the noise power is $\sigma^2 = -94$\,dBm.
The antenna gains are set at $5$\,dBi for what concerns the IRS/DF side, while the source and destination have isotropic antennas with gain $0$\,dBi.

\begin{figure}[t!]\vspace{-0.2cm}
\centering
    \begin{overpic}[width=0.92\columnwidth]{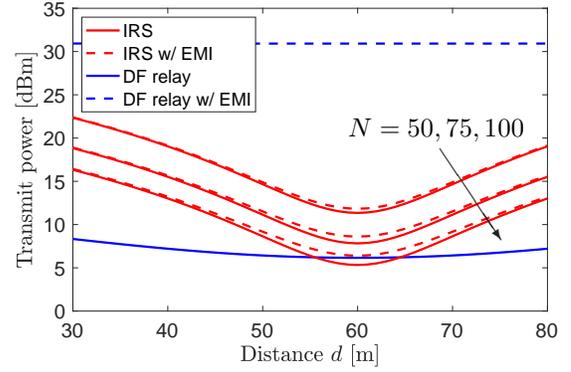}
    \put(58,38){\rotatebox{0}{{\textcolor{black}{$N=50,75,100$}}}}
    \put(73,36){\vector(2,-3){10}}
	\end{overpic}
	
	\caption{Transmit power required to obtain $\bar R = 6\,$bit/s/Hz with isotropic EMI, i.e., $\vect{R} = \vect{R}^{\rm iso}$. We assume that $\rho = \sigma^2_{\rm emi}/\sigma^2 = 25$~dB. The IRS is equipped with $N\in\{50,~75,~100\}$. The case without EMI (i.e., solid lines) is reported for comparisons.}
	\label{fig:fig3}\vspace{-0.5cm}
\end{figure}
\begin{figure}[t!]
\centering
	\includegraphics[width=0.92\columnwidth]{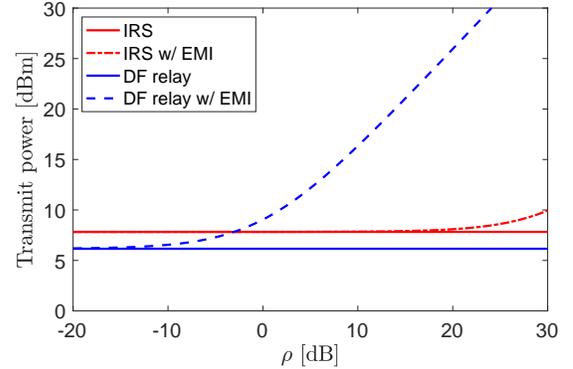} \vspace{-0.1cm}
	\caption{Transmit power required to obtain $\bar R = 6\,$bit/s/Hz in the setting of Fig.~\ref{fig:fig3} when $d=60$\,m. The IRS is equipped with $N= 75$ elements.}
	\label{fig:fig4}\vspace{-0.6cm}
\end{figure}
Comparisons are made in terms of transmit power required to achieve an information rate of $\bar R = 6$\,bit/s/Hz. The analysis focuses on two cases: without and with optimization against EMI. In former case, the transmit power with the IRS is obtained from~\eqref{eq:opt_power_ris} with $\phi_n = -\arg([{\bf h}_{\rm{sr}}]_{n} [{\bf h}_{\rm{rd}}]_{n})$, while it is computed according to~\eqref{eq:opt_power_df} with DF. In the latter case, the IRS is optimized according to the iterative algorithm in~\cite{torres2021} and then used to compute~\eqref{eq:opt_power_ris}, while Algorithm~\ref{alg:max-min-fairness} is used for the optimization of the DF relay. To proceed further, we define
\begin{align}
    \rho = \frac{\sigma^2_{\rm emi}}{\sigma^2}
\end{align}
as the ratio between the variances of EMI and thermal noise.
%according to Eq.~\eqref{eq:opt_power_ris},~\eqref{eq:opt_power_df} and Algorithm~\ref{alg:max-min-fairness}.
%Unless otherwise specified, the ratio $\rho = \sigma^2_{\rm emi}/\sigma^2$ is equal to $25$\,dB. 
%For a fair comparison in terms of complexity, we assume that only one RF-chain for both DF and IRS. Consequently, the DF-relay will use only one antenna. The benefit of having multiple antennas, and then multiple RF-chains, will be investigated in Sec.~\ref{sec:number_antennas}.
\subsection{Without Optimization Against EMI}
 Fig.~\ref{fig:fig3} plots the required transmit power (in dBm) as a function of the distance $d$. We assume isotropic EMI (i.e., $\vect{R} = \vect{R}^{\rm iso}$) with $\rho = \sigma^2_{\rm emi}/\sigma^2=25$\,dB. The solid lines refer to the case without EMI (i.e., $\sigma^2_{\rm emi} = 0$) and are reported as references. We see that EMI has a severe impact on DF relay-assisted communications. For all the considered cases, the transmit power with relaying is increased by more than $20$~dB. On the other hand, the impact of EMI on IRS-assisted communications is marginal, although it increases as $N$ grows. 
 While an IRS with large $N$ is needed to beat the DF relay without EMI (in line with \cite{bjorn2020irs}), our results clearly show that the IRS-assisted system is superior in the presence of EMI.

\begin{figure}[t!]\vspace{-0.25cm}
\centering
	\begin{overpic}[width=0.92\columnwidth]{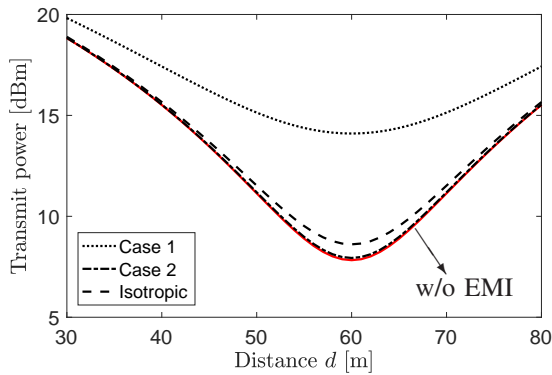}
	\put(70,24){\vector(2,-3){5}}
	\put(70,13){\rotatebox{0}{{{w/o EMI}}}}
	\end{overpic}
	\caption{Required transmit power to obtain a rate equal to $\bar R = 6\,$bit/s/Hz. The IRS is equipped with $N=75$. Different spatial distribution of the EMI are considered.}
	\label{fig:fig7}
\end{figure}

Fig.~\ref{fig:fig4} reports the required transmit power when $\rho = \sigma^2_{\rm emi}/\sigma^2$ varies and $d=60$~m. The EMI is again modelled as isotropic. In line with~\cite{bjorn2020irs}, DF performs better than an IRS with $N=75$ in the presence of weak EMI, i.e., for values of $\rho \le -3$~dB. However, the performance deteriorates quickly for $\rho \ge -3$~dB. On the other hand, an impact on IRS-assisted communications is observed only when $\rho \ge 20$~dB.
%
%\begin{figure}
%\centering
%	\begin{overpic}[width=0.92\columnwidth]{EMI_threeCases.eps}
%    
%	\put(22,3){\rotatebox{0}{(a)}}
%	\put(50,3){\rotatebox{0}{(b)}}
%	\put(84,3){\rotatebox{0}{(c)}}
%	
%	\put(55,13){\rotatebox{0}{$\sigma_{\theta}$}}
%	\put(3,22.5){\rotatebox{0}{$\sigma_{\theta}$}}
%	
%	\put(18,38){\rotatebox{0}{IRS/DF}}
%	\put(46,38){\rotatebox{0}{IRS/DF}}
%	\put(79,38){\rotatebox{0}{IRS/DF}}
%	
%	\end{overpic}
%	
%	\caption{The EMI arrives on the IRS/DF from: (a) the same angle as the source with $\sigma_\theta = 10^\circ$; (b) an angle orthogonal to that of the source with $\sigma_\theta = 10^\circ$ and  (c) all directions in front of the IRS/DF.}
%    \label{fig:simul_setup}
%\end{figure}

To quantify the impact of the EMI's spatial correlation properties, we now assume $f(\varphi,\theta)$ in~\eqref{eq:R1_expression} is of the form in~\cite[Eq.~(18)]{torres2021},
% \begin{align}\label{eq:non-isotropic_pdf}
% f(\varphi,\theta)  = \frac{c}{2 \pi \bar\sigma_{\varphi} \bar\sigma_{\theta}} e^{-\frac{(\varphi - \bar\varphi)^2}{2 \bar\sigma_{\varphi}^2}}e^{-\frac{(\theta - \bar\theta)^2}{2 \bar\sigma_{\theta}^2}} \cos (\theta)
% \end{align}
% where $c$ is a constant such that $\int_{0}^{\pi} \hspace{-1mm} \int_{-\pi}^{0} f_i(\varphi,\theta) d\varphi d \theta = 1$. This 
which represents a concentration of interfering plane waves arriving from the nominal angle pair $(\bar\theta, \bar\varphi)$ with a Gaussian angular distribution. We will compare two cases: 
\begin{enumerate}
    {\setlength\itemindent{13pt} \item[Case 1)] The nominal angles $(\bar\theta, \bar\varphi)$ are the same of the source signal, i.e., $\bar{\varphi}=\varphi_{\rm sr}$ and $\bar{\theta}=\theta_{\rm  sr}$;
    \item[Case 2)]  The nominal angles $(\bar\theta, \bar\varphi)$ are the same of the destination, i.e., $\bar{\varphi}=\varphi_{\rm  rd}$ and $\bar{\theta}=\theta_{\rm  rd}$.}
\end{enumerate}
In both cases, we assume that $\bar{\sigma}_{\theta}=\bar{\sigma}_{\varphi}=10^\circ$. Fig.~\ref{fig:fig7} plots the required transmit power with an IRS in the same setup as in Fig.~\ref{fig:fig3} but for two cases above. %We assume that $N=75$. 
The results show that the EMI is particularly severe when it impinges from the same angular direction as the source signal (Case 1). In Case 2, the IRS configuration will reflect the EMI in another direction than towards the destination. This demonstrates how the IRS is spatially filtering the EMI and mainly the part that resembles the desired signal will affect the destination.
%This means that That is to show that a wrong estimation of the EMI distribution at the IRS location site can dramatically affect its performance and then it should be carefully taken into account when building an IRS-assisted network.

\subsection{With Optimization Against EMI}
\begin{figure}[t!]\vspace{-0.25cm}
\centering
	    \begin{overpic}[width=0.92\columnwidth]{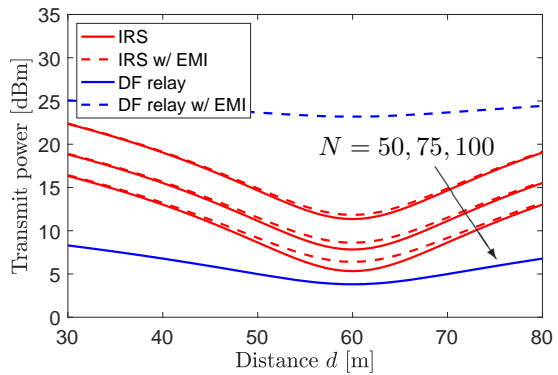}
    \put(54,36){\rotatebox{0}{{\textcolor{black}{$N=50,75,100$}}}}
    \put(73,34){\vector(2,-3){10}}
	\end{overpic}
% 	\caption{Transmit power required to obtain $\bar R = 6\,$bit/s/Hz in the setting of Fig.~\ref{fig:fig3} when the IRS and DF are both
% optimized against EMI.}\vspace{-0.5cm}
	\caption{Transmit power to obtain $\bar R = 6\,$bit/s/Hz in the setting of Fig.~\ref{fig:fig3} when the IRS and DF are optimized against EMI.}
	\label{fig:fig6}
\end{figure}

\begin{figure}[t!]\vspace{-0.5cm}
\centering
	\includegraphics[width=0.92\columnwidth]{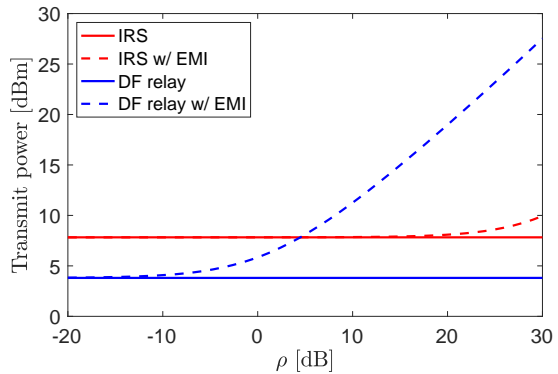}
	\caption{Required transmit power to obtain a rate equal to $\bar R = 6\,$bit/s/Hz in the setting of Fig.~\ref{fig:fig3} with $d=60$\,m and $N=75$. Both IRS and DF are optmized against EMI.}\vspace{-0.5cm}
	\label{fig:fig5}
\end{figure}

We now consider the case where both the IRS and DF relay are optimized against EMI.
Fig.~\ref{fig:fig6} plots the transmit power in the same setup as in Fig.~\ref{fig:fig3}. Compared to the previous results, we see that the IRS does not benefit much from the optimization. This is in line with the observations in~\cite{torres2021} where substantial gains are observed only when a much larger number of elements is used. The optimization of the DF relay provides more substantial benefits, but the it anyway loses its benefits compared to the IRS in the presence of EMI.
%Although the optimization of the relay provides some benefits, the impact of EMI remains relevant (compared to case where it is absent) and DF continues to perform much worse than IRS. 
This is confirmed by the results in Fig.~\ref{fig:fig5}, obtained in the same setting as in Fig.~\ref{fig:fig4} but with EMI-aware optimization. We see that the DF relay is now better than the IRS for $\rho \le 5$\,dB (not only for $\rho \le -3$\,dB). However, its performance continues to degrade fast for larger values of $\rho$.


\section{How many antennas are needed at the DF relay to beat an IRS under EMI?}
\label{sec:number_antennas}
So far, we have considered a single-antenna relay, {to match with the single RF chain that an IRS needs to select the optimal configuration \cite{bjornson2022reconfigurable}.} Next, we consider a relay with $M$ antennas and aim to determine how large $M$ must be to outperform an IRS in the presence of EMI. With a slight abuse of notation, we call $\vect{h}_{\text{sr}} \in \mathbb{C}^M$ the channel vector between the BS and DF, and $\vect{h}_{\text{rd}} \in \mathbb{C}^M$ the channel vector between the DF and destination. We still assume deterministic channels. In the first phase, after applying the combining vector $\vect{g}\in \mathbb{C}^M$,
the received signal  at the multi-antenna DF  is
\begin{align}\label{eq:DF_first_stage}
%\vect{g}^H\vect{y}_{\rm 1} = \vect{g}^H(\vect{h}_{\rm sr} \sqrt{p_1} s + \underbrace{\vect{n} + \vect{w}_{\rm 1}}_{\vect{z}})
 \vect{g}^H\left(\vect{h}_{\rm sr} \sqrt{p_1} s + \vect{n} + \vect{w}_{\rm 1}\right) = \vect{g}^H\left(\vect{h}_{\rm sr} \sqrt{p_1} s + \vect{z}\right)\!\!
\end{align}
where %$\vect{g}\in \mathbb{C}^M$ is the combining vector,
$\vect{z}=\vect{n} + \vect{w}_{\rm 1}\in \mathbb{C}^M\sim \CN({\bf 0}_M,\vect{C})$ with $\vect{C} = (\sigma_{\rm emi}^2 \vect{R} + \sigma^2 \vect{I}_M)$. The SNR in the first phase is, thus, given by\vspace{-1.5mm}
\begin{align}\label{eq:DF_first_stage_multiple}
	{\rm SNR_1} = p_1 \frac{|\vect{g}^H\vect{h}_{\rm sr}|^2}{\vect{g}^H{\bf C} \vect{g}} = p_1\alpha_1
\end{align}
where the effective channel gain in the first phase is now $\alpha_1=\frac{|\vect{g}^H\vect{h}_{\rm sr}|^2}{\vect{g}^H{\bf C} \vect{g}}$. The combiner $\vect{g}$ must be determined based on the available channel knowledge at the relay. If it knows only the  channel $\vect{h}_{\rm sr}$, then the optimal strategy is maximum ratio (MR) combining, which corresponds to $\vect{g} = \vect{h}_{\rm sr}$. As with the IRS, better performance can be achieved if the relay knows also the  EMI statistics, i.e., knowledge of $\sigma_{\rm emi}^2 \vect{R}$. In this case, we notice that the SNR in~\eqref{eq:DF_first_stage} is in the form of a generalized Rayleigh quotient and thus its maximum is achieved by the minimum mean-square-error (MMSE) combiner $\vect{g} = \vect{C}^{-1}\vect{h}_{\rm sr}$, which is obtained by whitening followed by MR combining.

\begin{figure}\vspace{-0.25cm}
     \centering
     \begin{subfigure}[b]{\columnwidth}
         \centering
         \begin{overpic}[width=0.92\columnwidth]{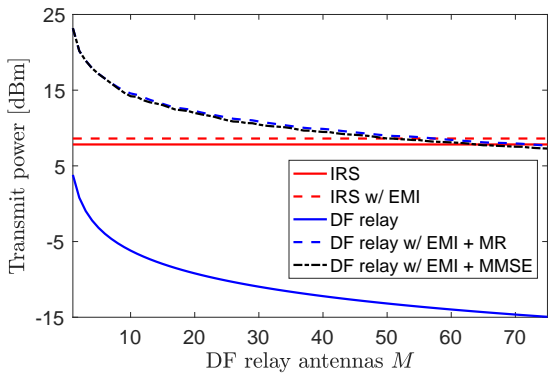}
        %  \put(40,48){\rotatebox{0}{{\textcolor{black}{MR}}}}
        %  \put(30,30){\rotatebox{0}{{\textcolor{black}{MMSE}}}}
        %  \put(34.5,43.5){\vector(1,1){5}}
        %  \put(34,42){\vector(0,-1){8}}
         \end{overpic}
         \caption{Isotropic propagation conditions}
         \label{fig:fig8A}
     \end{subfigure}
     \hfill
     \begin{subfigure}[b]{\columnwidth}
         \centering
         \begin{overpic}[width=0.92\columnwidth]{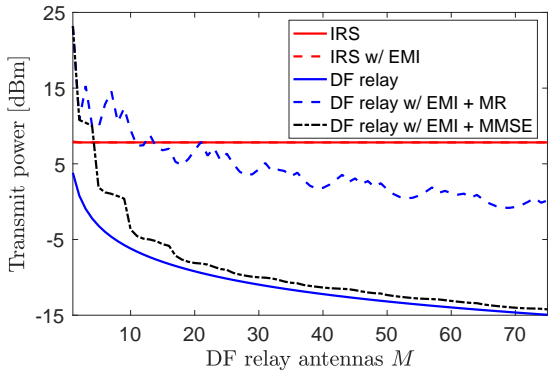}
        %  \put(59,21.5){\rotatebox{0}{{\textcolor{black}{MR}}}}
        %  \put(35.5,20){\rotatebox{0}{{\textcolor{black}{MMSE}}}}
        %  \put(57,31){\vector(1,-2){3}}
        %  \put(30,20){\vector(4,1){5}}
         \end{overpic}
         \caption{Propagation conditions: Case 2) $\bar{\varphi}=\varphi_{\rm  rd}$ and $\bar{\theta}=\theta_{\rm  rd}$.}
         \label{fig:fig8B}
     \end{subfigure}
        \caption{Required transmit power to obtain a rate equal to $\bar R = 6\,$bit/s/Hz in the setting of Fig.~\ref{fig:fig3} with $d=60$\,m . The IRS has $N=75$ elements and the DF uses MR and MMSE receivers.}
        \label{fig:fig8}\vspace{-0.5cm}
\end{figure}

In the second phase, the received signal is
\begin{align}
    y_2 = \vect{h}_{\rm rd}^T \vect{v}\sqrt{p_2}s + w_2
\end{align}
where $\vect{v}$ is the precoding vector and $w_2\sim \CN(0,\sigma^2)$ is the thermal noise. The SNR in the second phase is 
\begin{align}\label{eq:DF_second_stage_multiple}
    {\rm SNR}_2 = p_2 \frac{|\vect{h}_{\rm rd}^T \vect{v}|^2}{\sigma^2}=p_2\alpha_2
\end{align}
with $\alpha_2=\frac{|\vect{h}_{\rm rd}^T \vect{v}|^2}{\sigma^2}$ being the effective channel gain in the second phase. The SNR above is maximized by MR precoding, i.e., $\vect{v} = \vect{h}_{\rm rd}^* /\|\vect{h}_{\rm rd}\|$. The achievable rate is obtained from~\eqref{eq:Rdtfs} by plugging the effective channel gains $\alpha_1,\alpha_2$ from~\eqref{eq:DF_first_stage_multiple} and~\eqref{eq:DF_second_stage_multiple}, respectively. Once obtained, it can be used to run Algorithm~\ref{alg:max-min-fairness} for optimization of the relaying parameters.

\subsection{Performance Comparison}

Fig.~\ref{fig:fig8} plots the transmit power required by the DF relay as a function of $M$ with MR and MMSE combining. In Fig.~\ref{fig:fig8A}, we assume that the EMI is isotropic while in Fig.~\ref{fig:fig8B} we consider the propagation conditions of Case 2) from Fig.~\ref{fig:fig7}. In both cases, we assume $\rho=25$\,dB and $d=60$\,m. The DF relay is optimized against EMI by means of Algorithm~1 with $\alpha_1$ {in~\eqref{eq:DF_first_stage_multiple}} as obtained with MR or MMSE combining. Comparisons are made with an optimized IRS equipped with $N=75$. The performance of the DF relay without EMI is reported as a reference. In the case of isotropic EMI, the results of Fig.~\ref{fig:fig8A} show that MR and MMSE combining basically require the same power, which decreases as $M$ increases. Hence, having knowledge of $\sigma_{\rm emi}^2 \vect{R}$ does not provide much gain in the presence of isotropic EMI. In both cases, the DF relay needs at least $M=54$ antennas to match the performance of an IRS. %Compared to the case in which the DF relay operates in the absence of EMI, 
{The performance gap is large compared to the case without EMI.}
From Fig.~\ref{fig:fig8B}, we see that, when the EMI is non-isotropic but impinges on the IRS from a spatial direction that is sufficiently different from that of the source (Case 2), the DF relay can effectively suppress it by MMSE combining, and thus outperform the IRS. Particularly, we see that an extra  $M=5$ antennas (i.e., RF chains) are sufficient with MMSE combining to beat IRS. This number increases to $M=20$ with MR. In contrast to Fig.~\ref{fig:fig8A}, the large gap performance gap between MR and MMSE shows that, in the presence of spatially correlated EMI, having knowledge of its correlation matrix is highly valuable. We see that with $M\ge 20$, MMSE combining performs relatively close to the case without EMI. 
	
\section{Conclusions}

We provided a new comparison between IRS- and DF-assisted communications when EMI is present. Although the IRS aperture can capture much EMI, the many reflecting elements allow to partially mitigate its effect by means of spatial filtering; {mainly the EMI impinging from the same direction as the desired signal will be reflected to the destination.} On the contrary, the single-antenna DF relay is {very sensitive to} EMI, even if the optimal splitting of the two communication phases is considered. 
The analysis showed that IRS requires a much lower power to achieve a target rate. 
Multiple antennas (i.e., radio-frequency chains) are needed at the DF relay to be competitive against EMI. 
{If the EMI is spatially correlated, MMSE combining can be used to effectively suppress it.}

\bibliographystyle{IEEEtran}
% argument is your BibTeX string definitions and bibliography database(s)
\bibliography{refs}

% Generated by IEEEtran.bst, version: 1.14 (2015/08/26)
\begin{thebibliography}{1}
\providecommand{\url}[1]{#1}
\csname url@samestyle\endcsname
\providecommand{\newblock}{\relax}
\providecommand{\bibinfo}[2]{#2}
\providecommand{\BIBentrySTDinterwordspacing}{\spaceskip=0pt\relax}
\providecommand{\BIBentryALTinterwordstretchfactor}{4}
\providecommand{\BIBentryALTinterwordspacing}{\spaceskip=\fontdimen2\font plus
\BIBentryALTinterwordstretchfactor\fontdimen3\font minus
  \fontdimen4\font\relax}
\providecommand{\BIBforeignlanguage}[2]{{%
\expandafter\ifx\csname l@#1\endcsname\relax
\typeout{** WARNING: IEEEtran.bst: No hyphenation pattern has been}%
\typeout{** loaded for the language `#1'. Using the pattern for}%
\typeout{** the default language instead.}%
\else
\language=\csname l@#1\endcsname
\fi
#2}}
\providecommand{\BIBdecl}{\relax}
\BIBdecl

\bibitem{direnzo2020}
M.~Di~Renzo, A.~Zappone, M.~Debbah, M.-S. Alouini, C.~Yuen, J.~de~Rosny, and
  S.~Tretyakov, ``Smart radio environments empowered by reconfigurable
  intelligent surfaces: How it works, state of research, and the road ahead,''
  \emph{IEEE Journal on Selected Areas in Communications}, vol.~38, no.~11, pp.
  2450--2525, 2020.

\bibitem{wu2021}
Q.~Wu, S.~Zhang, B.~Zheng, C.~You, and R.~Zhang, ``Intelligent reflecting
  surface-aided wireless communications: A tutorial,'' \emph{IEEE Transactions
  on Communications}, vol.~69, no.~5, pp. 3313--3351, 2021.

\bibitem{bjornson2022reconfigurable}
E.~Bj{\"o}rnson, H.~Wymeersch, B.~Matthiesen, P.~Popovski, L.~Sanguinetti, and
  E.~de~Carvalho, ``Reconfigurable intelligent surfaces: A signal processing
  perspective with wireless applications,'' \emph{IEEE Signal Processing
  Magazine}, vol.~39, no.~2, pp. 135--158, 2022.

\bibitem{huang2019reconfigurable}
C.~Huang, A.~Zappone, G.~C. Alexandropoulos, M.~Debbah, and C.~Yuen,
  ``Reconfigurable intelligent surfaces for energy efficiency in wireless
  communication,'' \emph{IEEE Transactions on Wireless Communications},
  vol.~18, no.~8, pp. 4157--4170, 2019.

\bibitem{bjorn2020irs}
E.~Bj{\"o}rnson, {\"O}.~{\"O}zdogan, and E.~G. Larsson, ``Intelligent
  reflecting surface versus decode-and-forward: How large surfaces are needed
  to beat relaying?'' \emph{IEEE Wireless Communications Letters}, vol.~9,
  no.~2, pp. 244--248, 2020.

\bibitem{nossek2009}
M.~T. Ivrla\v{c} and J.~A. Nossek, ``Toward a circuit theory of
  communication,'' \emph{IEEE Transactions on Circuits and Systems I: Regular
  Papers}, vol.~57, no.~7, pp. 1663--1683, 2010.

\bibitem{torres2021}
A.~de~Jesus~Torres, L.~Sanguinetti, and E.~Bj{\"o}rnson, ``Electromagnetic
  interference in {RIS}-aided communications,'' \emph{IEEE Wireless
  Communications Letters}, pp. 1--1, 2021.

\bibitem{Bjornson21}
E.~Bj{\"o}rnson and L.~Sanguinetti, ``Rayleigh fading modeling and channel
  hardening for reconfigurable intelligent surfaces,'' \emph{IEEE Wireless
  Commun. Lett.}, vol.~10, no.~4, pp. 830--834, 2021.

\bibitem{3gpp}
3GPP, ``Further advancements for e-utra physical layer aspects (release 9),''
  \emph{Standard TS 36.814}, Mar. 2010.

\end{thebibliography}

\end{document}